\documentclass[3p,times,twocolumn]{elsarticle}
 \biboptions{comma,sort&compress}
 
\usepackage{graphicx}
\usepackage{ecrc}
\usepackage{here}

\volume{00}

\firstpage{1}

\journalname{Nuclear and Particle Physics Proceedings}

\runauth{S. Kawabata}

\jid{nppp}

\jnltitlelogo{Nuclear and Particle Physics Proceedings}

\usepackage{amssymb}

\usepackage[figuresright]{rotating}

\begin{document}

\begin{frontmatter}

\title{ Towards precise top mass measurement at LHC
 $^*$}
 \cortext[cor0]{Talk given at 18th International Conference in Quantum Chromodynamics (QCD 15,  30th anniversary),  29 june - 3 july 2015, Montpellier - FR}
 \author[label1,label2]{Sayaka Kawabata}
\ead{skawabata@seoultech.ac.kr}
\address[label1]{Department of Physics, Tohoku University, Sendai 980-8578, Japan}
\address[label2]{Institute of Convergence Fundamental Studies, Seoul National University of Science and Technology, Seoul 01811, Korea}


\pagestyle{myheadings}
\markright{ }
\begin{abstract}
The top quark mass plays an important role in a variety of discussions both within and beyond the Standard Model.
However, a precise determination of a theoretically well-defined top quark mass is still missing.
Towards a precise determination of a theoretically well-defined top quark mass at the LHC, we propose a method which uses lepton energy distribution and has a boost-invariant nature.
We investigate its experimental viability by performing a simulation analysis for $t\overline{t}$ production process and lepton$+$jets decay channel at the leading order.
We estimate several major uncertainties in the top mass determination with this method and they amount to $1.7$\,GeV with an integrated luminosity of $100$\,fb$^{-1}$ at $\sqrt{s}=14$\,TeV.
The uncertainties should be reduced by considering the next-to-leading order corrections to the method.
\end{abstract}

\end{frontmatter}
\section{Introduction}

Among the elementary particles in the Standard Model (SM), the top quark often plays a special role because of its large mass.
The top quark mass appears in a variety of important discussions, for example, electroweak precision tests~\cite{Baak:2012kk, Ciuchini:2013pca, Baak:2014ora}, vacuum stability~\cite{Degrassi:2012ry, Buttazzo:2013uya} and predictions beyond the SM, serving as a critical parameter.

Despite the above importance, a precise measurement of a theoretically well-defined top quark mass is still missing.
Top quark mass measurements have been performed at the Tevatron and the Large Hadron Collider (LHC) using event kinematics of $t\overline{t}$ final states.
These measurements achieved a top mass determination with a precision well below $1$\,GeV~\cite{ATLAS:2014wva, Tevatron:2014cka, Khachatryan:2015hba}.
However, they extracted the top quark mass by fitting event kinematics including jet momenta with Monte-Carlo (MC) simulations, and therefore, the measured mass depends on a hadronization model in the MC simulators.
In this sense, the measured mass is referred to as ``MC mass" and should be distinguished from well-defined masses in perturbation theory.
Since it is difficult to evaluate accurately uncertainties of hadronization models, relations between the MC mass and theoretically well-defined masses are not known so far.

Theoretically well-defined top masses, like the $\overline{\rm MS}$ mass and the pole mass, have also been extracted from $t\overline{t}$ production cross section measurements.
The $\overline{\rm MS}$ mass measurement by the D0 collaboration yields $m_t^{\overline{\rm MS}}=160.0\,^{+5.1}_{-4.5}$\,GeV~\cite{Abazov:2011pta} using the approximate NNLO calculation of Ref.~\cite{Moch:2008qy}.
The pole mass measurements by the ATLAS and the CMS collaborations yield $m_t^{\rm pole}=172.9\,^{+2.5}_{-2.6}$\,GeV~\cite{Aad:2014kva} and $m_t^{\rm pole}=173.6\,^{+1.7}_{-1.8}$\,GeV~\cite{CMS:2015uoa}, respectively.
In addition, another approach to measure the pole mass was recently applied by the ATLAS collaboration, which yields $m_t^{\rm pole}=173.7\,^{+2.3}_{-2.1}$\,GeV~\cite{Aad:2015waa}.
The approach utilizes $t\overline{t}+1$\,jet differential distribution, improving the sensitivity to the top mass compared to using $t\overline{t}$ cross section~\cite{Alioli:2013mxa}.
These measurements, however, still have large errors compared to the MC mass measurements.
In order to achieve a high precision of less than $1$\,GeV, a drastic change of approaches will be required.

Another important issue of the top quark mass determination is an intrinsic limit on accuracy related to the pole mass.
From the point of view of perturbative expansion, the quark pole mass is not suitable because its perturbative series exhibit bad convergence properties.
This reflects the fact that the top quark has a color and thus a physical on-shell quark cannot exists.
It is known that its perturbative prediction cannot avoid the ambiguity of the order of $\Lambda_{\rm QCD}$~\cite{Bigi:1993zi,Bigi:1994em,Beneke:1998ui}.
This difficulty can be circumvented by using so-called short-distance masses, such as the potential-subtracted (PS) mass~\cite{Beneke:1998rk}, the 1S mass~\cite{Hoang:1999zc} and the $\overline{\rm MS}$ mass.
Among them, the $\overline{\rm MS}$ mass is commonly used and its good convergence properties in perturbative expansions are confirmed in various observables.
The relation between the pole mass and the $\overline{\rm MS}$ mass of the top quark is known up to four-loop order in perturbative QCD~\cite{Marquard:2015qpa}.


The aim of this study is to determine a theoretically well-defined top quark mass, especially the $\overline{\rm MS}$ mass, at the LHC.
In order to achieve this aim, we propose the ``weight function method."
The method was first proposed in Refs.~\cite{Kawabata:2011gz,Kawabata:2013fta} as a new method to reconstruct properties of a parent particle.
We applied the method to the top quark mass reconstruction at the LHC in Ref.~\cite{Kawabataa:2014osa}.
This paper follows these references.
The method has two distinct features.
First, it uses only lepton energy distribution as an observable.
Second, it has a boost-invariant nature.
We will explain the method in more detail in the next section.
To investigate its experimental viability, we perform a simulation analysis of the top mass reconstruction, considering detector acceptance, event selection cuts and background contributions.
The analysis is performed at the leading order (LO).
Theoretical corrections, such as higher-order QCD corrections and finite-width effects of the top quark, will be taken into account in our future works.

The paper is organized as follows.
In section~\ref{sec:WFM}, we outline our method for the top mass reconstruction.
In section~\ref{sec:simulation}, we perform a simulation analysis at LO and present the results.
Summary and future prospects of this study are given in section~\ref{sec:summary}.

\section{Weight Function Method}
\label{sec:WFM}

In this section, we first outline the weight function method, and then try to explain why the method works.

Suppose that a particle decays into many particles including a lepton and we are interested in measuring the mass of the parent particle.
We assume the parent particle is unpolarized with respect to its boost direction.
In the method, we use a characteristic weight function $W(E_\ell,m)$.
The weight function is given by
\begin{equation}
	\!\!\!\!\!\!\!\!\!W(E_\ell,m)=\!\int \!dE \left.\mathcal{D}_0(E;m)\frac{1}{EE_\ell} \,({\rm odd~fn.~of~}\rho)\right|_{e^{\rho}=E_\ell/E}\,,
	\label{eq:WF}
\end{equation}
where $\mathcal{D}_0(E;m)$ is the normalized lepton energy distribution in the rest frame of a parent particle with a mass parameter of $m$.
The distribution $\mathcal{D}_0(E;m)$ can be calculated theoretically if we know the decay process of the parent particle.
We can choose an arbitrary odd function in Eq.~(\ref{eq:WF}).

Using the weight function $W(E_\ell,m)$ and a (normalized) lepton energy distribution $D(E_\ell)$, we define a weighted integral $I(m)$ by
\begin{equation}
	I(m) \equiv \int dE_\ell \,D(E_\ell) \,W(E_\ell,m)\,.
	\label{eq:Im}
\end{equation}
Then, $I(m)$ satisfies
\begin{equation}
	I(m=m^{\rm true}) = 0\,,
	\label{eq:zero}
\end{equation}
where $m^{\rm true}$ is the true value of the mass of the parent particle.
Thus, the mass of the parent particle can be reconstructed as the zero of $I(m)$ with a lepton energy distribution as an observable.

Note that the theoretical prediction required in this method (in the definition of a weight function) is the lepton energy distribution $\mathcal{D}_0$ in the rest frame of a parent particle.
In contrast, the experimental observable compared with it is a lepton energy distribution $D(E_\ell)$ in any Lorentz frame.
That is, we can obtain the mass independently of velocity distribution of the parent particle.

\begin{figure}[htb]
		\includegraphics[width=.7\hsize]{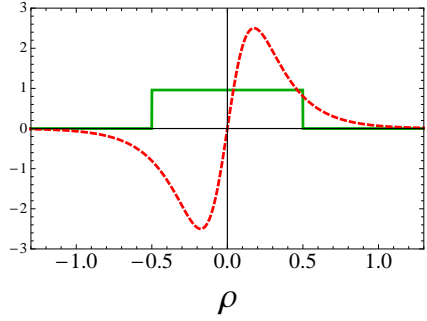}
		\caption{Lepton energy distribution $\mathcal{D}(E_\ell;\,\beta)$ in a boosted frame (the solid line) and $e^{\,\rho}W(E_\ell,m^{\rm true})$ (the dashed line) in terms of $\rho=\log (E_\ell/E_0)$. $\beta$ is chosen such that $y=0.5$\,, which is the edge point of $\mathcal{D}(E_\ell;\,\beta)$, and $E_0$ is taken to be $1$.}
		\label{fig:RhoDist}
\end{figure}

We can easily see why this works in two-body decay case.
In this case, the lepton energy distribution in the rest frame of the parent particle is a delta function: $\mathcal{D}_0(E;m)=\delta (E-E_0)$, where $E_0$ is expressed with the parent particle mass $m$.
The weight function [Eq.~(\ref{eq:WF})] for the two-body decay becomes
\begin{equation}
	W(E_\ell,m) = \left.\frac{1}{E_0 E_\ell} ({\rm odd~fn.~of~}\rho)\right|_{e^{\rho}=E_\ell/E_0}\,.
\end{equation}
Lepton energy distribution in a boosted frame with a velocity $\beta$ is obtained as
\begin{equation}
	\!\!\!\!\!\!\!\mathcal{D} (E_\ell\,;\beta) ~=~ \frac{1}{2E_0 \sinh y} ~\theta \left( e^{-y} E_0 \,\leq\, E_\ell \,\leq\, e^y E_0 \right)\,,
	\label{eq:LepEneDist_2BodyDecay}
\end{equation}
where $y$ is defined by $e^{2y} = (1+ \beta)/(1- \beta)$ and $\theta \,(condition)$ is $1$ if the $condition$ is true and otherwise $0$.
With the weight function and the lepton energy distribution written as a function of $\rho$, Eq.~(\ref{eq:Im}) becomes
\begin{eqnarray}
	&&\hspace{-1.1cm}\int dE_\ell \,\mathcal{D}(E_\ell\,;\beta) \,W(E_\ell,m) \nonumber \\
		&&\hspace{-1cm}\propto \int d\rho~\theta \left( -y \,\leq\, \rho \,\leq\, y \right)\times ({\rm odd~fn.~of~}\rho) = 0 \,,
		\label{eq:RhoDist}
\end{eqnarray}
if $m=m^{\rm true}$.
Figure~\ref{fig:RhoDist} illustrates this situation.
The weighted integral $I(m)$ becomes zero as a result of integrating a product of an even function and an odd function.
Note that the velocity of the parent particle $\beta$ affects the width of the rectangle distribution $\mathcal{D}(E_\ell;\,\beta)$ in the figure.
However, the distribution is an even function of $\rho$, irrelevantly to the width, and therefore, Eq.~(\ref{eq:RhoDist}) holds in a boost-invariant way.
This property holds even if the velocity of the parent particle has a distribution.

In the case of many-body decay, the explanation is not as simple as in the two-body decay case.
Although we do not prove that for many-body decay in this paper, you can easily follow the proof in Refs.~\cite{Kawabata:2011gz,Kawabata:2013fta}.

The weight function method uses only lepton, which is a comparatively clear object in hadron colliders and free from hadronization.
It means that the method is suitable to determine theoretically well-defined top quark masses.
Moreover, it has a boost-invariant nature, and owing to this, the required theoretical prediction is only the lepton distribution in the rest frame of a parent particle.
Therefore, it is (basically) independent of the production process of the parent particle.
There is a disadvantage:
the method requires, in principle, the whole distribution of lepton energy.
In real experiments, the lepton distribution is distorted by various experimental effects, for example, event selection cuts and background contributions.
Consequently, a reconstructed mass with such distorted distribution in the weight function method deviates from the true mass value.
We overcame this disadvantage in the case of top mass reconstruction, by modifying the method (see the next section)~\cite{Kawabataa:2014osa}.

\section{Simulation Analysis at LO}
\label{sec:simulation}

Top quarks produced in $t\overline{t}$ pair production at the LHC are almost unpolarized with respect to their boost directions.
Parity-violating weak interactions induce a small polarization, whose size at the LHC is at sub-percent level in the SM prediction~\cite{Bernreuther:2006vg,Bernreuther:2013aga}.
We ignore this effect in the analysis, and will include it as a small correction in our future analysis if necessary.

\begin{figure}[htb]
		\includegraphics[width=.75\hsize]{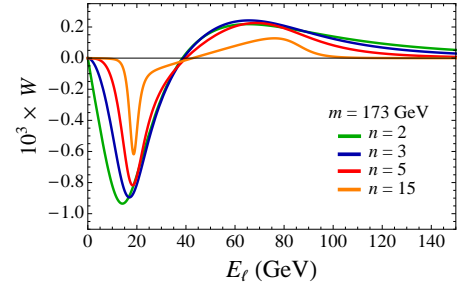}
		\caption{Weight functions $W(E_\ell,m)$ used in the analysis with $m=173$\,GeV.
		The odd function in Eq.~(\ref{eq:WF}) takes the form of Eq.~(\ref{eq:oddfunc}) for $n=2,\,3,\,5$ and $15$.
}
		\label{fig:WF}
\end{figure}

We apply the weight function method to the top mass reconstruction using the lepton $\ell$ in the top quark decay $t \rightarrow b \ell \nu$.
The weight function defined in Eq.~(\ref{eq:WF}) is obtained with the lepton energy distribution $\mathcal{D}_0(E;m)$ in the rest frame of the top quark with a mass parameter $m$.
We compute the distribution $\mathcal{D}_0(E;m)$ at LO.
There is a degree of freedom to choose an odd function in the definition of the weight function.
We choose the following odd functions for this analysis:
\begin{equation}
	({\rm odd~fn.~of~}\rho)\,=\,n\tanh (n\rho)/\!\cosh(n\rho)\,,
	\label{eq:oddfunc}
\end{equation}
with $n=2,\,3,\,5$ and $15$.
The resultant weight functions are shown in Fig.~\ref{fig:WF}.

We generate $t\overline{t}$ signal events decaying into lepton$+$jets at $\sqrt{s}=14$\,TeV at the LHC.
For the background events, we consider other $t\overline{t}$, $W+$jets, $Wb\overline{b}+$jets and single-top production events.
Other $t\overline{t}$ events include all the decay channels of $t\overline{t}$ except the signal channel.
Both signal and background events are generated at LO.
For details of the simulation setup, see Ref.~\cite{Kawabataa:2014osa}.

\begin{table*}[t]
\caption{Estimates of uncertainties in GeV from several sources in the top mass determination at LO. The weight function used in this evaluation corresponds to $n=2$. The signal and background statistical errors correspond to those with an integrated luminosity of $100$\,fb$^{-1}$.}
\label{tab:uncertainties}
\vspace{0.5mm}
\begin{tabular*}{\textwidth}{@{\extracolsep{\fill}}ccccc}
\hline
		~Signal stat. error~ & ~~~~~Fac. scale~~~~~ & ~~~~~~~PDF~~~~~~~ & ~Jet energy scale~ & ~Background stat. error~ \\ \hline
		$0.4$ & $+1.5$/$-1.4$ & $0.6$ & $+0.2$/$-0.0$ & $0.4$\\
\hline
\end{tabular*}
\end{table*}

\begin{figure}[htb]
		\includegraphics[width=.75\hsize]{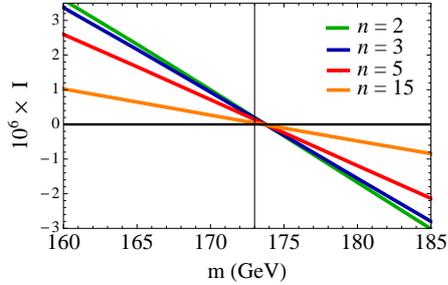}
		\caption{Weighted integrals $I(m)$ with the parton-level lepton distribution and the weight functions corresponding to $n=2,\,3,\,5$ and $15$. The input value of the top quark mass is $173$\,GeV.
}
		\label{fig:ImParton}
\end{figure}

We first confirm validity of the weight function method at the parton level.
Fig.~\ref{fig:ImParton} shows weighted integrals $I(m)$ with the parton-level lepton energy distribution of the generated signal events.
The weight functions used correspond to $n=2,\,3,\,5$ and $15$ in Eq.~(\ref{eq:oddfunc}) and
the input value of the top quark mass to the simulation events is $173$\,GeV.
In Fig.~\ref{fig:ImParton}, the deviations of the zeros of $I(m)$ from the true mass value are around $+0.7$\,GeV.
In this analysis, estimated statistical errors due to the limited MC simulation events are around $0.4$\,GeV.
In addition, the top quark width effect contributes to the deviation and is estimated to be $+0.34$\,GeV in our analysis.
Therefore, we confirm that the reconstructed masses reproduce the true mass within the estimated errors of the MC simulation.

In real experiments, the lepton distribution is distorted by various experimental effects.
In Ref.~\cite{Kawabataa:2014osa}, we examined each of various factors of anticipated experimental effects.
The most serious effect is caused by event selection cuts concerning leptons.
We deal with these effects in the following manner:
after applying event selection cuts,
(1) we estimate background contributions and subtract them, and then
(2) restore the lepton distribution to that of signal events at the parton level.
(3) Constructing the weighted integral $I(m)$ with the restored lepton distribution,
(4) we obtain the zero of $I(m)$ as the reconstructed mass.
In the procedure (2), we take a strategy of compensating for the loss caused by lepton cuts using MC events. 
Since the effects of lepton cuts are top-mass dependent, the compensating MC events also depend on the top quark mass (we call the mass $m_t^{\rm c}$). 
Therefore, we should solve $I(m=m_t^{\rm rec})=0$ with a consistency condition $m_t^{\rm rec}=m_t^{\rm c}$ in the procedure (4).

\begin{figure}[htb]
		\includegraphics[width=.75\hsize]{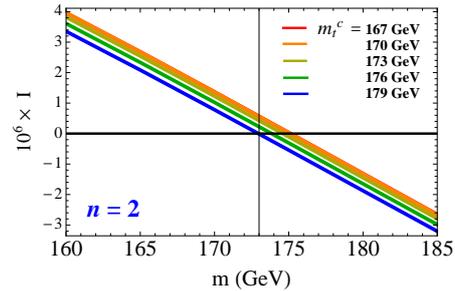}
		\caption{Weighted integrals $I(m)$ with various $m_t^{\rm c}$ after all the cuts. The weight function used corresponds to $n=2$. The input value of the top quark mass is $173$\,GeV.}
		\label{fig:Im_AftCuts}
\end{figure}

Fig.~\ref{fig:Im_AftCuts} shows the weighted integrals $I(m)$ using the compensated lepton distribution with $m_t^{\rm c}=167,\,170,\,173,\,176$ and $179$\,GeV.
The input value of the top quark mass is $173$\,GeV and the weight function used in this figure corresponds to $n=2$.
The zeros of $I(m)$ are close to the true mass value despite the compensated events with various masses $m_t^{\rm c}$.
The reconstructed mass from these weighted integrals is $174.1$\,GeV.
Taking into account the MC statistical error $+1.0$/$-1.1$\,GeV and the estimated shift due to the top width $+0.34$\,GeV in the analysis, the deviation from the true value $+1.1$\,GeV is consistent with these effects.
We vary the input top quark mass to the simulation events from $167$\,GeV to $179$\,GeV and perform the same top mass reconstruction.
Results show that reconstructed masses agree with the input masses, within the errors of the simulation analysis.

We estimate sensitivity of the method in the top quark mass determination.
In addition to signal and background statistical errors, we estimate uncertainties due to factorization scale dependence, parton distribution function (PDF) and jet energy scale.
The results are shown in Table~\ref{tab:uncertainties}.
The estimated signal statistical error is $0.4$\,GeV for lepton$+$jets channel with an integrated luminosity of $100$\,fb$^{-1}$.
The uncertainties related to factorization scale dependence and choice of PDF are involved due to the compensated simulation events.
Thus, these uncertainties are expected to be reduced by including the next-to-leading order (NLO) corrections to the production process of the top quark in a simulation event generator.
The uncertainty due to jet energy scale contributes indirectly through event selection cuts concerning jets.
The background statistical error accompanies the subtraction of background contributions.
Combining the uncertainties in Table~\ref{tab:uncertainties} amounts to about $1.7$\,GeV.

\section{Summary and Prospects}
\label{sec:summary}

We proposed a new method to measure the top quark mass at the LHC.
The method has features of using only lepton energy distribution and being (in principle) independent of velocity distribution of the top quark.
We performed a simulation analysis of the top mass reconstruction with the method at LO, using $t\overline{t}$ production events and its lepton$+$jets decay channel.
As a result, we confirmed that the method works even with lepton distribution distorted by experimental effects.
We estimated uncertainties from several major sources in the top mass measurement and they amount to $1.7$\,GeV corresponding to an integrated luminosity of $100$\,fb$^{-1}$ at $\sqrt{s}=14$\,TeV.

This analysis is still the first step of our project towards a precise determination of the top quark mass.
The top quark pole mass can be obtained with the method, computing the lepton energy distribution in the top quark rest frame used in the weight functions at NLO and NNLO in the on-shell scheme.
In addition, we will include NLO corrections to the top quark production process in the MC simulator used to generate the compensated events.
This should reduce uncertainties due to factorization scale dependence, which was the dominant source of uncertainties in the LO analysis.
Finite-width effects of the top quark should also be considered as a small correction to the method.
Including these theoretical corrections will improve the sensitivity of the top mass determination and will be steps towards the determination of the $\overline{\rm MS}$ mass with this method.

\section*{Acknowledgements}
This study is performed in collaboration with Y. Shimizu, Y. Sumino and H. Yokoya, and I wish to thank all of them.
This research was supported by Basic Science Research Program through the National Research Foundation of Korea (NRF) funded by the Ministry of Science, ICT and Future Planning (Grant No. NRF-2014R1A2A1A11052687).

\end{document}